\documentclass[referee]{aa}  

\usepackage{graphicx}
\usepackage{cuted,multicol}
\usepackage{txfonts}

\usepackage{amsmath}

\newcommand\wpewce{\omega_{pe}/\Omega_{ce}}
\newcommand\Wce{\Omega_{ce}}
\newcommand\wpe{\omega_{pe}}
\newcommand\wpet{$\omega_{pe}^{-1}$}
\usepackage[normalem]{ulem}

\begin{document}

        \title{Excitation of extraordinary modes inside the source of Saturn's kilometric radiation}
        
        
        \author{Hao Ning
                \inst{1,2}
                \and Yao Chen\inst{1,2}
                \and Chuanyang Li\inst{1,2}
                \and Shengyi Ye\inst{3}
                \and Alexey Kuznetsov\inst{4}
                \and Siyuan Wu\inst{3}
        }
        
        \institute{Institute of Frontier and Interdisciplinary Science, Shandong University, Qingdao, Shandong, 266237, People's Republic of China\\
                \email{yaochen@sdu.edu.cn}
                \and
                Institute of Space Sciences, Shandong University, Shandong, 264209, People's Republic of China\\
                \and
                Department of Earth and Space Sciences, Southern University of Science and Technology, Shenzhen, Guangdong, People’s Republic of China\\
                \and
                Institute of Solar-Terrestrial Physics, Irkutsk, 664033, Russia\\
        }
        
        \date{Received ; accepted }
        
        
        \abstract
        {The electron cyclotron maser instability (ECMI)  of extraordinary mode waves was investigated with the parameters observed in Saturn's kilometric radiation (SKR) sources. Previous studies employed simplified dispersion relations, and did not consider the excitation of the relativistic (R) mode. This mode is introduced by considering the relativistic effect in plasmas consisting of both cold and hot electrons. Using particle-in-cell simulations, we investigated the excitation of R and X modes based on the measured data. Using the reported value of the density ratio of energetic to total electrons $n_e/n_0=24\%$, the most unstable mode is the R mode. The escaping X-mode emissions are amplified only if the energetic electrons are dominant with $n_e/n_0 \ge 90\%$. For these cases, only the X mode is excited and the R mode disappears due to its strong coupling. The results are well in line with the linear kinetic theory of ECMI. The properties of both the R and X modes are consistent with the observed SKR emissions. This raises questions about the nature of the measured electric field fluctuations within ``presumed'' SKR sources. The study provides new insights into the ECMI process relevant to SKR emission mechanisms.}

        \keywords{Planets and satellites: gaseous planets -- 
                Radio continuum: planetary systems -- 
                Masers -- Waves -- Plasmas -- 
                Methods: numerical
        }

        \titlerunning{ECMI of SKR sources}
        \maketitle

\section{Introduction} \label{sec:intro}

Saturn’s kilometric radiations (SKRs) represent intense radio emissions originating from the kronian auroral region, considered to be the counterpart of the auroral kilometric radiations (AKRs) on Earth \citep[see, e.g.,][]{1980Sci...209.1238K,1981Natur.292..731K,2004SSRv..114..395G,2005Natur.433..722K, 2009JGRA..11410212L}.
The radiations are observed in the kilohertz to megahertz frequency range, and the radiated powers of $10^8$--$10^{10}$ watts. 
Furthermore, SKRs are closely related to Saturn's magnetosphere dynamics \citep[][]{2011pre7.conf....1J,2013JGRA..118.7019K}, and have been extensively studied.

Saturn’s kilometric radiations are often strongly polarized with narrow beaming angles and high intensity \citep{2008JGRA..113.7201L,2018Sci...362.2027L}, indicating coherent emissions generated by electron cyclotron maser instability (ECMI) \citep{1979ApJ...230..621W, 1985SSRv...41..215W, 2006A&ARv..13..229T,2013SSRv..178..695B}. The instability excites escaping modes at gyro-frequencies ($\sim \Wce$) via a gyro-resonant wave-particle interaction, driven by electrons with anisotropic velocity distribution functions \citep[VDFs; see][for recent studies]{2021A&A...651A.118N, 2021ApJ...920L..40N}. The ECMI is now widely applied to planetary and stellar radio emissions.

The Cassini spacecraft measures the radio waves with the Radio and Plasma Wave Science \citep[RPWS, ][]{2004SSRv..114..395G} instrument and the electrons with the Cassini Plasma Spectrometer Electron Spectrometer \citep[CAPS/ELS, ][]{1998GMS...102..257L, 2004SSRv..114....1Y} in Saturn's magnetosphere. 
Since 2010, researchers have reported several crossings of SKR sources by Cassini, identified with the cutoff frequency of the spectrum close to or slightly lower than the local $\Wce$ \citep[e.g.,][]{2011pre7.conf...75K, 2010GeoRL..3719105M, 2010GeoRL..3712104L,  2011JGRA..11612222M, 2018Sci...362.2027L}. 

As has previously been reported, SKR sources are characterized by a low ratio of plasma oscillation frequency to electron gyro-frequency ($\wpewce <0.1$), in accordance with the ECMI conditions \citep[$\wpewce<1$,][]{1979ApJ...230..621W}. The emissions mainly propagate quasi-perpendicularly to the background magnetic field with linear polarization near the sources, consistent with the X-mode properties. The energy conversion rate from electrons to radiations is $\sim1\%$ \citep{2011JGRA..116.4212L}. The abundances of energetic and cold electrons are comparable with each other \citep{2011JGRA..116.5203S}.

The measurement of electron VDFs is crucial to the study of the ECMI process. In the first encounter of the SKR source by Cassini in 2008, \cite{2010GeoRL..3719105M} found that the cold and hot electrons can be fitted with a Kappa and a Dory-Guest-Harri \citep[DGH, ][]{1965PhRvL..14..131D} distribution, respectively. They further calculated the corresponding ECMI growth rates of X-mode emissions with the linear kinetic theory, and suggested that the ECMI process can efficiently amplify this escaping X mode below $\Wce$, along directions perpendicular to the magnetic field. \cite{2011JGRA..11612222M} investigated the ECMI process of another SKR source using a similar method and obtained the same results. They also suggested that O-mode and Z-mode waves can be generated. 

However, these studies on the ECMI of SKRs followed \cite{2007JGRA..112.7211M} and employed the dispersion relations of pure relativistic energetic electrons. They estimated the correction to the dispersion relation with a scale factor when calculating the growth rates of the X and Z modes, rather than solving the dispersion matrix to calculate the real frequency and the growth rate in a self-consistent manner. As reported by \cite{2010GeoRL..3719105M} and \cite{2011JGRA..11612222M}, the density ratio of the energetic to the total electrons was fitted as $n_e/n_0 = 20\%$--24\%. With this range of $n_e/n_0$ (and other comparable values), one obtains very different dispersion relations. For waves perpendicular to the magnetic field, the gyro-resonance condition is simplified as $\omega=n\Wce/\gamma< n\Wce$. 
When adopting the cold-plasma dispersion relation deduced from the magneto-ionic theory, the ECMI process cannot amplify the fundamental X mode with the cutoff frequency ($\omega_{\text{X}}$) being larger than $\Wce$. For pure energetic electrons, the X-mode cutoff is less than $\Wce$ and thus can grow via ECMI \citep[e.g.,][]{1983PlPh...25..217W,1984GeoRL..11..143P,1985ApJ...291..160W,1986JPlPh..35..187R}.

For plasmas consisting of both the cold and energetic electrons, another important mode may emerge. This mode, first deduced by \cite{1984JGR....89.8957P}, appears in the range of $\Wce/\gamma <\omega<\Wce$. It is relevant to the relativistic effect, and is therefore called the relativistic (R) mode \citep{1985JGR....90.9675S, 1986JGR....91.3152S}. Our present study shows that the R mode becomes the most unstable mode, and is probably relevant to SKR. However, it is trapped and cannot escape. This mode has not been considered in studies on SKR sources, to the best of our knowledge.

According to the earlier reports on the crossings of SKR sources by Cassini, both cold and hot components of electrons exist with comparable densities \citep{2011JGRA..116.5203S}. This raises  questions concerning the applicability of the oversimplified dispersion relations adopted by \cite{2010GeoRL..3719105M} and \cite{2011JGRA..11612222M}, which did not consider the existence of the R mode. 
For this study, we conducted fully kinetic electromagnetic particle-in-cell (PIC) simulations with the electron VDFs and plasma parameters reported by \cite{2011JGRA..11612222M} to investigate the extraordinary modes (including R, X, and Z) and their coupling within SKR sources. We conducted a parametric study on $n_e/n_0$, and applied the kinetic linear theory to verify and explain the PIC simulations.

\section{Numerical method and parameter setting} \label{sec:method} 

As reported by \cite{2011JGRA..11612222M}, the frequency ratio was $\wpe/\Wce=0.046$, and the VDFs of both the cold and hot electrons were written as 
\begin{equation}\label{eq1}
        f_0 = \frac{1}{(2\pi)^{3/2}v_0^3}\exp (-\frac{u^2}{2v_0^2})
\end{equation} 
and
\begin{equation} \label{eq2}
        f_e=\frac{1}{j!(2\pi)^3/2v_\perp^2 v_\parallel}\left( \frac{u_\perp^2}{2 v_\perp^2}\right)^j \exp \left(-\frac{u_\perp^2}{2v_\perp^2}-\frac{u_\parallel^2}{2v_\parallel^2}\right ),
\end{equation}
respectively, where $j = 11$, $u$ represents the momentum per mass, and $v_\parallel~(= 0.11c)$ and $v_\perp~(=0.038c)$ represent the thermal velocity of the DGH distribution along parallel and perpendicular directions. The density ratio of the energetic to the total electrons was fitted to be $n_e/n_0=24\%$. Considering the uncertainties of the source measurements, we conducted a parametric study of $n_e/n_0$. The models and parameter settings are presented below.

\subsection{PIC simulations} \label{sec:picmodel}

We employed the Vector-PIC (vPIC) code developed by the Los Alamos National Laboratory (LANL). The code employs a second-order, explicit, leapfrog algorithm to update the positions and the velocities of charged particles, along with a full Maxwell description of the electric and magnetic fields via a second-order finite-difference time-domain solver \citep{2008PhPl...15e5703B,2009JPhCS.180a2055B}. 

We performed 2D3V (two spatial dimensions with three velocity components) simulations in the $xOz$ plane with periodical boundary conditions and a uniform background magnetic field ($\vec B_0 =B_0 \hat e_z$). The simulation domain was set to $L_x=L_z=2048\Delta$, the grid length to $\Delta=0.019~\lambda_{DE}$, and the time step to $\Delta t = 0.0097~ \wpe^{-1}$. 

\begin{figure*}[htbp!]
        \centering
        \includegraphics[width=16cm]{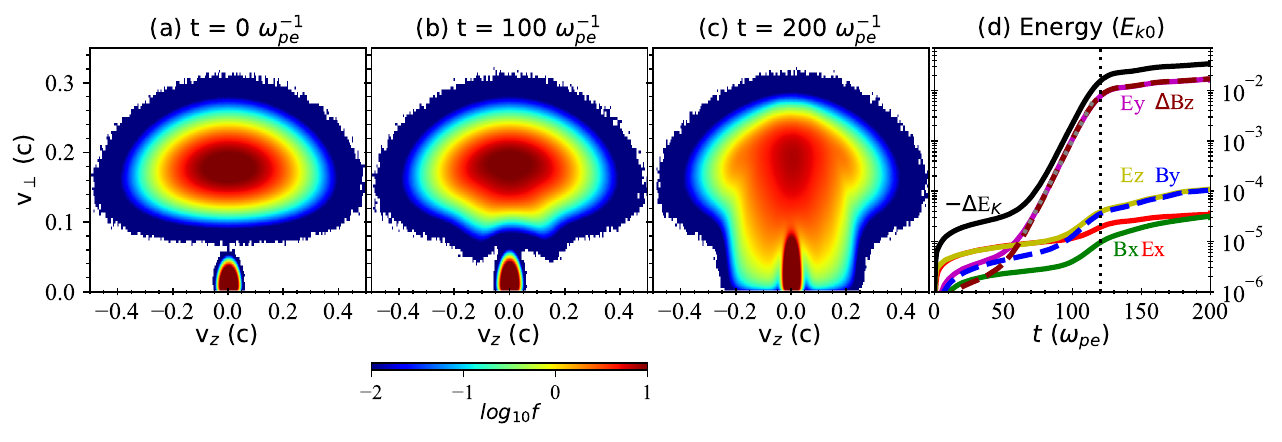}
        \caption{Evolution of electron VDFs and energy profiles of the simulation with $n_e /n_0=24\%$. Panels (a--c): Snapshots of the VDFs at the start (a, $t=0~\omega_{pe}^{-1}$), middle (b, 100~\wpet), and end (c, 200~\wpet) of the simulation. Panel (d): Temporal variations of energies of the six fluctuated field components. The black line represents the decline of the electron kinetic energy ($-\Delta E_k$). The energies are normalized to the initial kinetic energy of the total electrons. The gray dashed line represents the exponential fitting of the energy profiles. The vertical dotted line in (d) indicates the separate time (120~\wpet) of the linear stage and the saturation stage.}
        \label{fig:1}
\end{figure*}

Two components of electrons were included. Their distribution functions are described by Eqs. (\ref{eq1}) and (\ref{eq2}), and the center of the VDF of energetic electrons is at $v_\perp\sim0.18~c$ (Fig.~\ref{fig:1}(a)). The thermal velocity of the low-energy electrons was set to $v_0 =0.01c$ ($\sim50$ eV). The charge neutrality was maintained by including thermal protons with a realistic proton-to-electron mass ratio of 1836 and the same temperature (with the thermal velocity $v_p = v_0/\sqrt{1836}$). We included 1000 macroparticles per cell for each component of the electrons and 500 for the protons. 

\begin{figure*}[htb!]
        \centering
        \includegraphics[width=16cm]{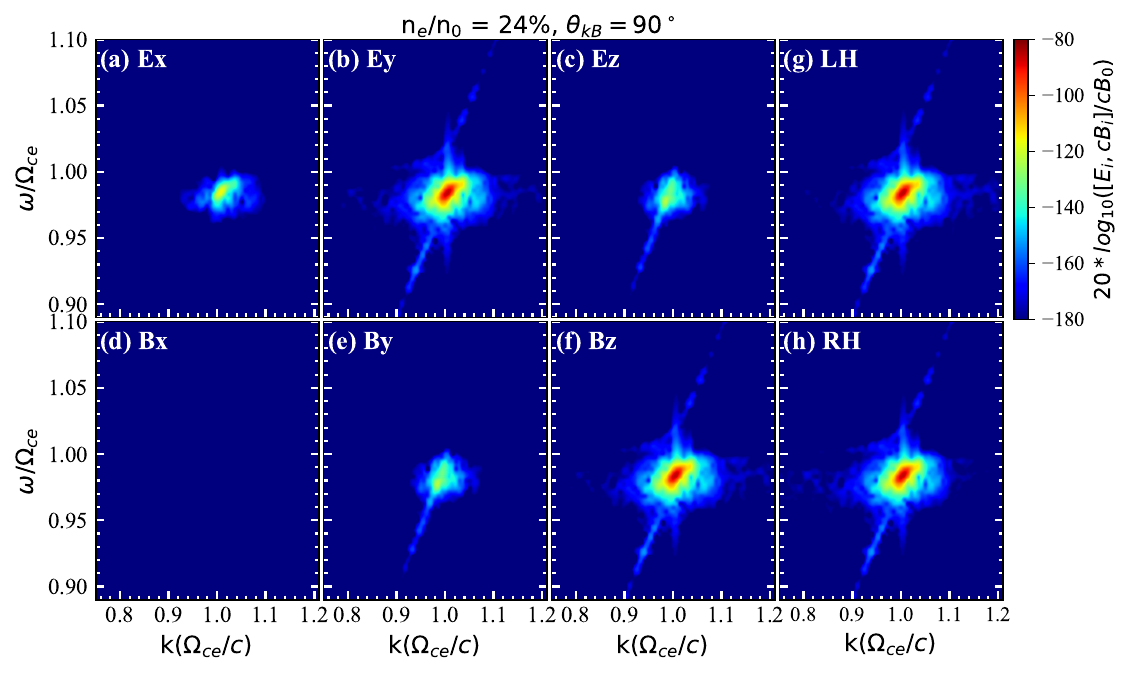}
        \caption{Wave dispersion diagrams of electric and magnetic fields with $n_e/n_0 = 24\%$ and $\theta_{kB}=90^\circ$, during the interval of [100, 200]~\wpet. The six field components are provided in panels (a--f) and the separated LH and RH components in panels (g--h).}
        \label{fig:2}
\end{figure*}

\subsection{Linear analysis of ECMI}

Following the analysis of \cite{1984JGR....89.8957P}, we employed the $\delta$ function below to describe the two components of electrons,
\begin{equation} \label{eq:delta}
        f=(1-\frac{n_e}{n_0})\delta(u_\perp)\delta(u_\parallel) + \frac{n_e}{n_0}\delta(u_\perp-u_r)\delta(u_\parallel),
\end{equation}
where $u_r$ refers to the momentum of the energetic electrons. In accordance with the PIC simulation, we assumed the waves propagate in the $xOz$ plane with $\vec B_0 = B_z \hat e_z$. For simplicity, only the perpendicular propagating waves with $k_\parallel = 0$ were studied.

With the VDF presented above, we derived the elements of the plasma dispersion tensor (Eq.~(\ref{eq:tensor})) and obtained the frequencies and the growth rates of the ECMI modes by numerically solving their determinant. We analyzed the dispersion relations, the ECMI growth rates, and the polarization of the X and R modes. The results are presented in the appendix.

\section{PIC simulations} \label{sec:picresults}

We  present the PIC simulations with the reported SKR source parameters as the reference case with ($n_e/n_0=24\%$) to show the excitation and properties of extraordinary modes (R, X, Z), followed by the parametric study on $n_e/n_0$ to explore its effect.

\subsection{Results with measured parameters ($n_e/n_0=0.24$) } \label{sec:picsource}

According to the energy profiles presented in Fig.~\ref{fig:1}(d), the wave excitation and further evolution can be divided into two stages, the linear (0--120~\wpet) stage and the saturation (120--200~\wpet) stage. In the linear stage, the energies of $\Delta B_z~(=B_z-B_0)$ and $E_y$ rise rapidly, reaching the level of $10^{-2}~E_{k0}$ ($E_{k0}$ is the initial kinetic energy of total electrons). The electron kinetic energy ($E_k$) declines in accordance with the growth of the ECMI modes. The energies of other field components are much weaker, being $\sim10^{-5}~E_{k0}$. In the saturation stage, the energies of all the fields rise gradually. 
In total, $\sim$3.4\% of $E_{k0}$ was converted to wave energies.

Figure~\ref{fig:1}(a--c) presents the evolution of the electron VDFs. The two components are initially separated. Within 100--200~\wpet, the shell-like hot electrons diffuse in the phase space, later developing a butterfly-like morphology. Meanwhile, the cool 
Maxwellian electrons are significantly energized along the perpendicular direction.

We performed a fast Fourier transform (FFT) analysis on each field component. In Fig.~\ref{fig:2}(a--f), we show the $\omega - k$ dispersion diagrams for perpendicular-propagating waves with $\theta_{kB}=90^\circ$ (the angle between $\vec k$ and $\vec B_0$), along which the ECMI waves are the most amplified. The strongest wave appears in the $E_y$ and $B_z$ components, with frequencies of $\sim0.98~\Wce$, in accordance with the dispersion relation of the R mode as shown in the Appendix. This mode can be identified as the transverse electromagnetic wave with significant $E_y$ and $B_z$ components and an insignificant $E_x$ component. The property of the simulated R mode is in line with the linear kinetic theory (see Section~\ref{sec:theory}).

To separate the left-hand (LH) and right-hand (RH) polarized components of the R mode, we processed ($E_x$, $E_y$) with the following equation: 
\begin{equation} 
        E_{LH, RH}=\frac{1}{\sqrt{2}}(E_x\pm iE_y).  \label{eq:LR} 
\end{equation}
The results are presented in Fig.~\ref{fig:2}(g--h). The intensities of  the LH and RH components are comparable to each other, indicating a strong linear polarization. 

According to the energy profiles in Fig.~\ref{fig:1}(d), the total energy of the R mode is $\sim$0.034~$E_{k0}$, and its growth rate can be fitted as $\Gamma = 5.57\times 10^{-3}~\Wce$. The fundamental O mode is slightly amplified in $E_z$ and $B_y$. Its total energy is weaker than the R mode by approximately two orders of magnitude.

In this case, only the R mode can be amplified, with no significant growth of X. The results can be clearly understood with the linear kinetic theory. 
As shown in the Appendix, the R mode is the most unstable at $k\sim\Wce/c$ with a maximal growth rate of $0.006~\Wce$, and linearly polarized in $E_y$ with $E_x/iE_y\sim0$ (Fig.~\ref{fig:A1}(a--c)). The growth rates of the X mode remain at 0 with its cutoff being larger than $\Wce$.

\subsection{Effect of $n_e/n_0$} \label{sec:picn}

\begin{figure*}[!b]
        \centering
        \includegraphics[width=12cm]{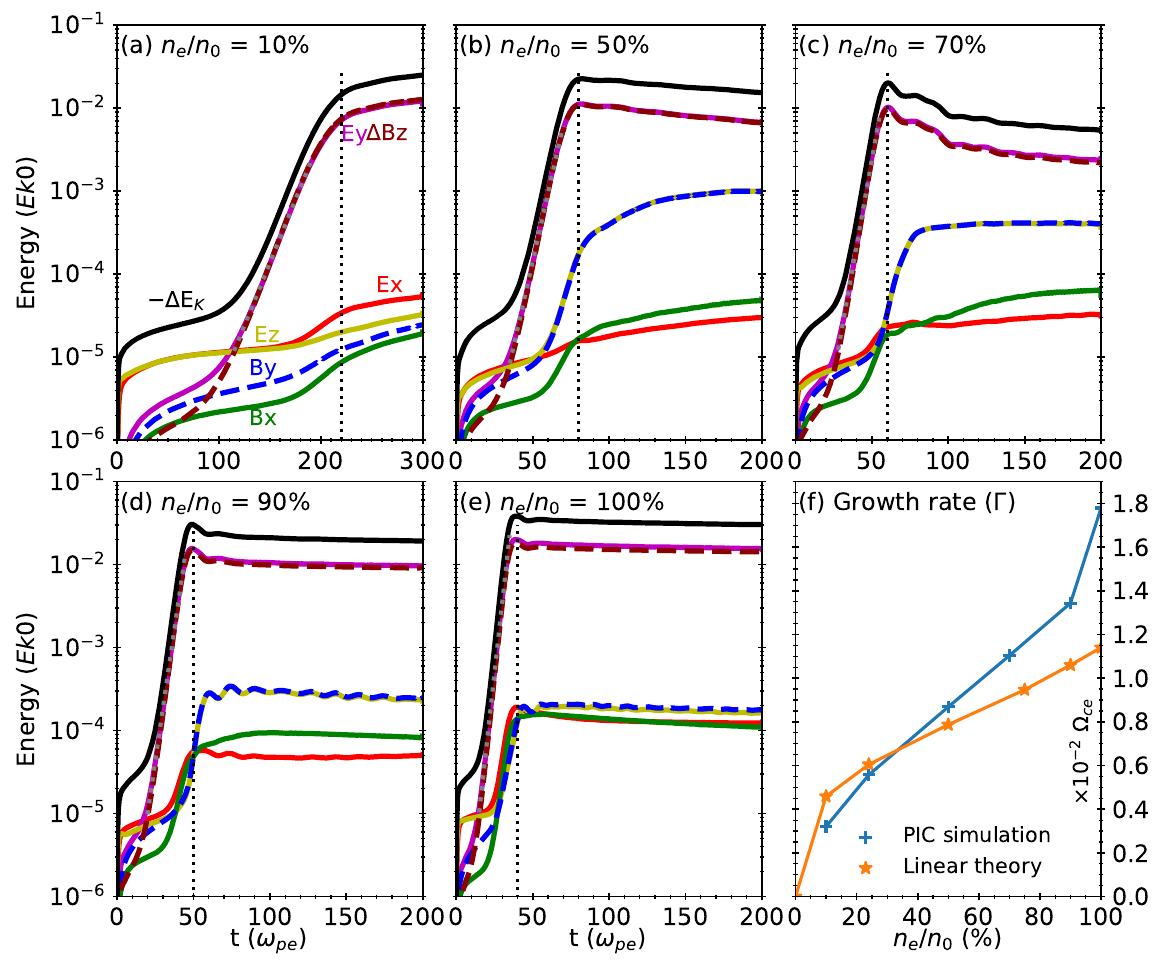}
        \caption{Energy profiles and growth rates for different simulation cases. Panels (a--e): Energy profiles of the six fluctuated field components for the cases with $n_e/n_0=$ 0.1, 0.5, 0.7, 0.9, and 1.0. The energies are normalized to the respective initial kinetic energy of total electrons of each case. Panel (f) presents the growth rates with different density ratios obtained by the PIC simulations (blue) and the linear kinetic theory (orange).
        }
        \label{fig:3}
\end{figure*}

\begin{figure*}[htb]
        \centering
        \includegraphics[width=12cm]{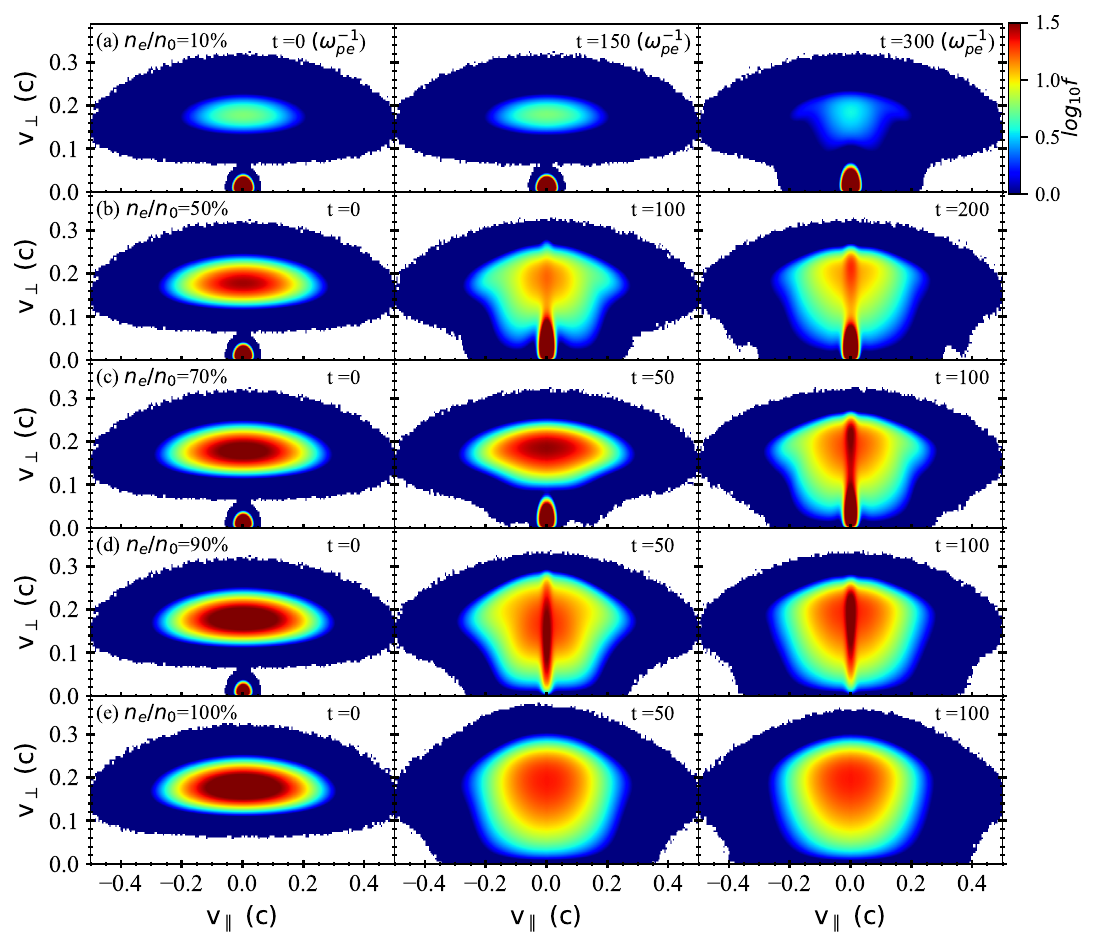}
        \caption{Snapshots of VDFs of simulations with $n_e/n_0=$ 10\% (a), 50\% (b), 70\% (c), 90\% (d), and 100\% (e) at moments representative of the initial (left), the transition (middle), and the saturation (right) stages.}
        \label{fig:4}
\end{figure*}
Considering the uncertainties of the electron measurement of Cassini CAPS/ELS, we varied the values of $n_e/n_0$ from 10\% to 100\% to investigate the conditions for X-mode excitations. For cases with a $n_e/n_0>10\%$, the simulations lasted for 200~\wpet, and for the $n_e/n_0 = 10\%$ case, the simulation lasted for 300~\wpet.

In Fig.~\ref{fig:3}(a--e), we plotted the energy profiles of each  case. As introduced in Section~\ref{sec:picsource}, we divided the whole ECMI process into two stages, the linear stage and the saturation stage. For each case we observed a sharp rise of $E_y$ in the linear stage. The energy conversion rates from the electron kinetic energy to the waves is $\sim10^{-2}$. With a higher $n_e/n_0$, the wave excitation is more impulsive, with a short linear stage. For cases with $n_e/n_0 \ge 50\%$, we observed a gradual decline of the field energies and $-\Delta E_k~(=E_{k0}-E_k)$, suggesting further damping of the amplified waves. 
Figure~\ref{fig:4} displays the evolution of the VDFs in different cases. With a larger $n_e/n_0$, the VDF of the hot electrons diffuses faster, and the cold electrons are  more efficiently accelerated. The VDFs do not change much after 50--100~\wpet.

\begin{figure*}[htb!]
        \centering
        \includegraphics[width=0.8\textwidth]{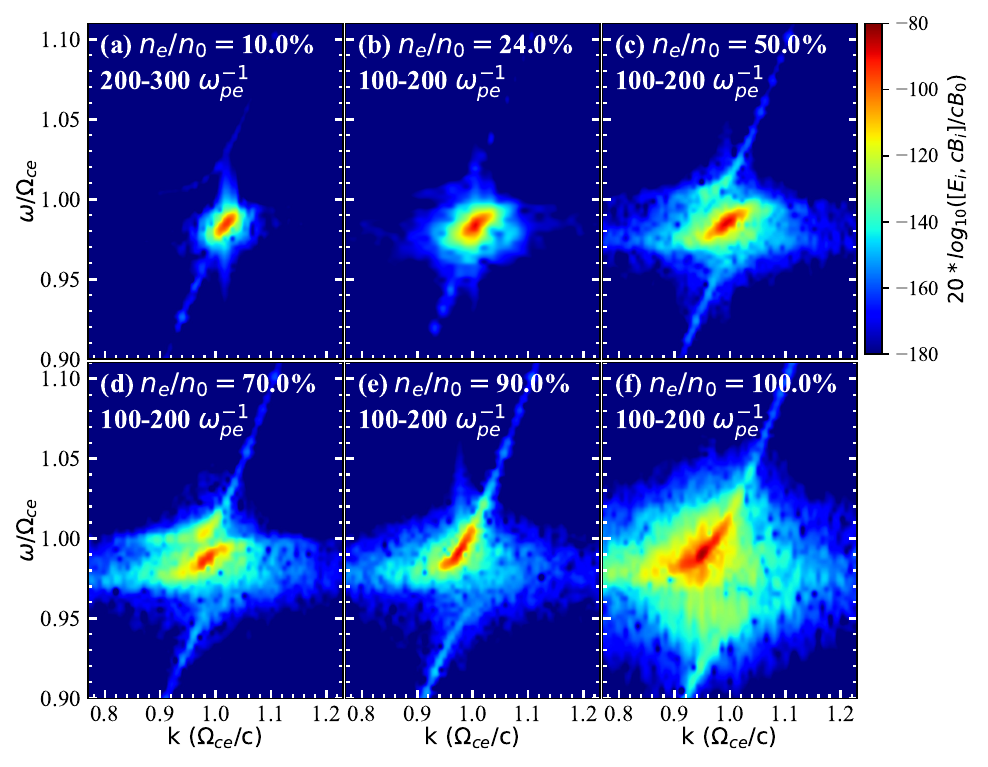}
        \caption{$\omega-k$ dispersion diagrams of $E_y$ at $\theta_{kB}=90^\circ$ during the saturation stage for different cases. }
        \label{fig:5}
\end{figure*}

\begin{figure*}[htb!]
        \centering
        \includegraphics[width=0.8\textwidth]{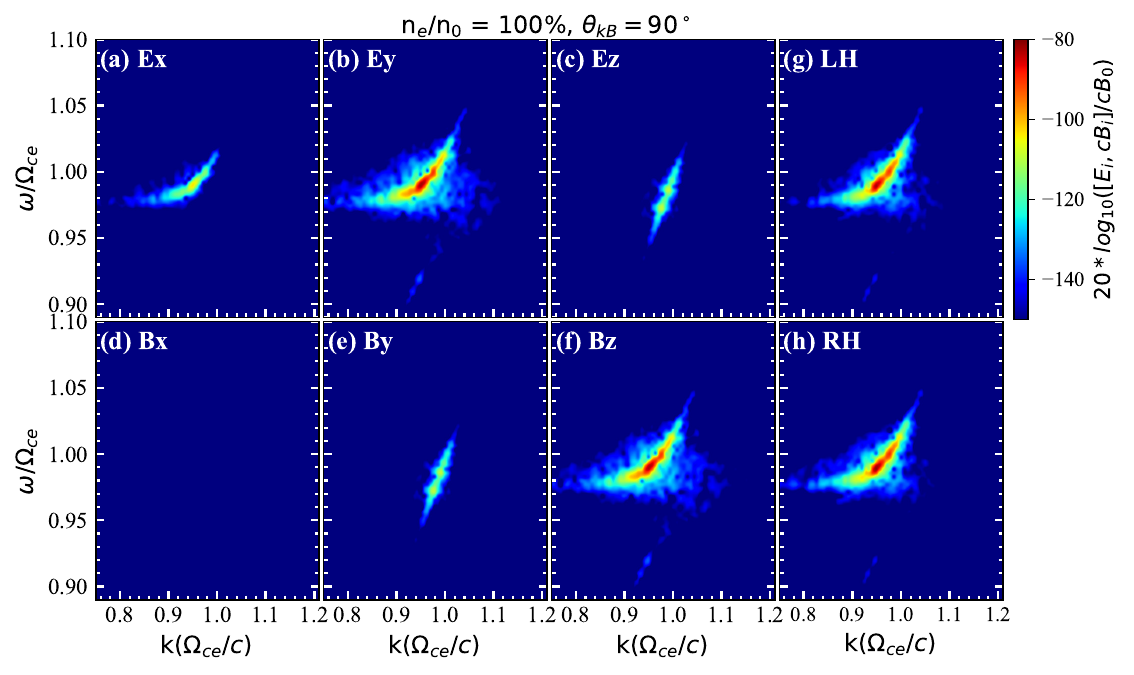}
        \caption{Wave dispersion diagrams of the six field components (a--f) and the separated LH and RH components (g--h) with $n_e/n_0=100\%$ and $\theta_{kB}=90^\circ$ during [100, 200]~\wpet. }
        \label{fig:6}
\end{figure*}

For each case, we performed a FFT analysis of the fluctuated fields to show the amplified wave modes (Fig.~\ref{fig:5}). The waves in $E_x$ and $E_z$ are not shown here since they are very weak. For cases with $n_e/n_0\le70\%$, the R mode is the most intensive, while no significant X mode appears. With increasing $n_e/n_0$, the R-mode dispersion curve becomes flatter and closer to that of the X mode. With $n_e/n_0 \ge 90\%$, the two dispersion curves merge, indicating their strong coupling; this leads to the significant excitation of the X mode with a lower cutoff frequency of $\omega_{\text{X}}\sim 0.97~\Wce$.

We plotted the growth rates ($\Gamma$) of the simulated R mode (or the X mode for the cases with $n_e/n_0=90$ and 100\%) by fitting the energy curves (Fig.~\ref{fig:3}(f)). The values of $\Gamma$ rise from 0.0032 to 0.018~$\Wce$ with $n_e/n_0$ from 10\% to 100\%. With the linear kinetic theory (see Fig.~\ref{fig:A2} of the Appendix), we obtained comparable growth rates for the R (for the cases with $0<n_e/n_0<100\%$) and the X mode (for the case with $n_e/n_0=100\%$), which rise from $4.59\times10^{-3}$ to $1.14\times10^{-2}~\Wce$.

According to our PIC simulations, the X mode is excited only if the energetic component is dominant ($n_e/n_0\ge 90\%$). The characteristics of the X mode in the case of $n_e/n_0=100\%$, including the growth rate, polarization, and wave fields (see Figs.~\ref{fig:3}(e), \ref{fig:4}(e), and \ref{fig:6}), are consistent with the linear kinetic theory (Fig.~\ref{fig:A1}(d--f)). In this case, the X mode becomes saturated within the first 50~\wpet, mainly in $E_y$ and $B_z$, and much less so in $E_x$. The energy conversion rate from energetic electrons to the X mode is $\sim$3\%, and the growth rate is $\sim$0.018~$\Wce$, similar to that obtained by the linear kinetic theory ($\sim0.014~\Wce$; see Fig.~\ref{fig:A1}(e)). According to Fig.~\ref{fig:6}(g--h), the amplified X-mode waves are linearly polarized with comparable LH and RH intensities. 

With the linear kinetic theory (see Section~\ref{sec:ratio_theory}), for the case with $n_e/n_0 = 90\%$, the growth rate of the X mode is 0. This is inconsistent with the corresponding PIC simulation presented above. This can be explained by the evolution of the VDFs (Fig.~\ref{fig:4}(d)). Almost all of the cool electrons are energized in the simulation within the first 50~\wpet, and their VDFs overlap with the hot component. This makes it difficult to distinguish the two components in the phase space, leading to an equivalent case with $n_e/n_0 = 100\%$. This explains why the simulation with $n_e/n_0 = 90\%$ is in line with that of $n_e/n_0 = 100\%$. 
In other cases with less energetic electrons, the two components of the VDF are still separated until the end of the simulation, and the X mode cannot grow. 
Further discussion is provided in the following section.

\section{Conclusion and discussion}\label{sec:conclusion}

The present study investigates the excitation of the R and X modes based on the in situ observation of  SKR sources reported by \cite{2011JGRA..11612222M}. Both PIC simulations and linear theory analysis were performed, and the results are in line with each other. With the given set of parameters, the trapped R mode is amplified as the dominant mode, while the escaping X mode can not grow. The growth rate of the R mode increases with a $n_e/n_0$ of 10\%--70\%. The X-mode emissions can only be amplified if the electrons are dominated by hot components ($n_e/n_0 = 90\%$ and 100\%) with the cutoff frequency less than $\Wce$ and the growth rate $\sim0.018~\Wce$. According to our PIC simulations, both the X and R modes can be characterized with similar frequencies ($\sim\Wce$),  polarization (linearly polarized in $E_y$), growth rates ($10^{-2}~\Wce$), and energy conversion rates (1\%--3\%). The properties can be well explained using the linear kinetic theory.

As the kinetic theory suggests, the X mode can be amplified only with $n_e/n_0=100\%$. In other cases with $0<n_e/n_0\le90\%$, the R mode is the most unstable and the X mode cannot grow. In our simulation with $n_e/n_0=90\%$, we observed significant declines of wave energies after the linear stage, with the cold electrons significantly accelerated at 50~\wpet. We propose a multi-step process to explain the X-mode excitation in this case (and its inconsistency with the linear theory): the R-mode waves are first amplified via ECMI, then interact strongly with the cool electrons and become damped soon via the gyro-resonant absorption. This results in the bulk acceleration of the minor cool electrons, which then become the ``hot'' component. This leads to an effective ``$n_e/n_0=100\%$'' case, and thus the strong coupling of the R and X modes and the strong excitation of the X mode. This means that the non-escaping R mode plays a crucial role in the dynamic evolution of electron VDFs and the excitation of the X mode. 

However, with the actual measured parameters, we obtained significant excitation of the trapped R mode rather than the escaping X-mode radiation. This does not explain SKR observed from a distance. We propose two possibilities to explain this problem:

1. The measured electron VDFs are not the direct driver of ECMI due to the poor resolution. The CAPS/ELS instrument measures the electrons of different pitch angles using eight detectors, which rotate for approximately 3 min to cover $2\pi$ sr of the sky. The temporal resolution is much larger than the timescales of the wave growth ($\sim100$\wpet). It is hard to measure electrons right before the excitation of ECMI since the VDFs diffuse rapidly at a timescale of 50--100~\wpet. Therefore, the ECMI of SKRs may not be induced by the measured electron VDFs.

2. The in situ observation cannot distinguish the R and X modes in the source region. According to our simulations and the linear theory analysis, both the R and X modes are identified as transverse electromagnetic waves with $\omega\sim\Wce$, linearly polarized in $E_y$. These properties are consistent with the observed SKR emissions. With the Cassini in situ measurements of the electric field only, it is difficult to distinguish the two wave modes. In other words, it is possible that the enhancement of the fluctuating electric fields observed by Cassini near or within the sources is actually the trapped R mode, rather than the generally believed escaping X mode.

Saturn's kilometric radiation has long been considered to be the analogy of the AKR from the geo-magnetosphere. Similar problems were present when applying the ECMI mechanism to AKRs. It is believed that  AKRs are generated from low-density cavities with no cold electron population. According to the PIC simulations of the finite source cavity \citep{1986JGR....9113569P, 1989JGR....94..129P, 2002JGRA..107.1437P}, the generated X-mode radiations could escape from the cavities via mode conversion at the boundaries. Although the presence of such cavity structures has not been conclusively identified in SKR sources, this could be a possible explanation for SKR. 

In the present simulations, we employed periodical boundary conditions to model the excitation of the wave modes, assuming the electrons are uniformly distributed in a large spatial scale. No replenishment of the electron free energy was included in this model. We will conduct simulations with various setups of the source configurations to examine their effect on the generation process of SKR .

The ECMI growth rates of the X and Z modes in SKR sources have been studied by  \cite{2010GeoRL..3719105M} and \cite{2011JGRA..11612222M}. As we pointed out in Section~\ref{sec:intro}, previous studies employed oversimplified dispersion relations with inappropriate assumptions and did not consider the R mode. In this paper, we performed PIC simulations to study the excitation of the three extraordinary modes (R, X, and Z), focusing mainly on the R mode and its effect on the excitation of the X mode. We also performed linear analysis of the plasma kinetic theory to obtain the dispersion relations and wave growths. The results are in agreement with the PIC simulations.

The R mode is dominant in most cases with both the cold and hot electrons, reaching  energies of $\sim 0.03~E_{k0}$. This mode is trapped between plasma layers delimited by the two electron gyro-frequencies ($\Wce/\gamma ,~\Wce$), yet it may be converted to escaping emissions when propagating in inhomogeneous plasmas. In addition, the R mode can be absorbed by electrons, leading to the acceleration and heating of plasmas, which, in turn, can modify the dispersion relations. Further simulations should investigate the heating and acceleration process due to the R mode and its coupling with the escaping X mode.

\begin{acknowledgements}
        This study is supported by NSFC grants (11790303 (11790300), 12203031, and 12103029), the China Postdoctoral Science Foundation (2022TQ0189), and the Natural Science Foundation of Shandong Province (ZR2021QA033). We thank the National Supercomputer Centers in Tianjin and the Beijing Super Cloud Computing Center (BSCC, URL: http://www.blsc.cn/) for providing HPC resources, and LANL for the open-source VPIC code. 
\end{acknowledgements}
        
        %
        %

\begin{appendix} 
        \section{Linear analysis of plasma kinetic theory}
        \subsection{Dispersion tensor}
        
        We followed \cite{1984JGR....89.8957P} to derive the dielectric tensors of plasmas with the VDF given by the $\delta$ function (Eq.~(\ref{eq:delta})). For perpendicular propagating waves, the elements of plasma dispersion tensors were simplified as

        \begin{equation}
                \begin{aligned}
                        &\varsigma_{xx} =  1-r^\prime\frac{\wpe^2}{\omega^2-1}- r\sum_{n=-\infty}^{\infty} \frac{n^2\wpe^2}{\xi^2\omega b}\left(2\xi J_n J_n^\prime - G_r\frac{\omega J_n^2}{b}\right) \\                                                
                        &\varsigma_{xy} =  ir^\prime\frac{\wpe^2}{(\omega^2-1)\omega}\\
                        &~~~~~~~~+ir\sum_{n=-\infty}^{\infty}  \frac{n\wpe^2}{\omega\xi b}\left[J_n J_n^\prime +\xi (J_n J_n^{\prime\prime} + J_n^{\prime2})-G_r\frac{\omega J_n J_n^\prime}{b}\right]=-\varsigma_{yx} \\
                        & \varsigma_{yy} = 1 - \frac{k^2 c^2}{\omega^2}-r^\prime\frac{\wpe^2}{\omega^2-1}\\
                        &~~~~~~~~- r\sum_{n=-\infty}^{\infty} \frac{\wpe^2}{\omega b} \left[ 2 J_n^{\prime 2} +2\xi J_n^\prime J_n^{\prime\prime} - G_r \frac{\omega J_n^{\prime 2}}{b}  \right] \\
                        &\varsigma_{xz}=\varsigma_{zx}=\varsigma_{yz}=\varsigma_{zy}=0\\
                        &  \varsigma_{zz} = 1-\frac{k^2 c^2}{\omega^2} - r^\prime\frac{\wpe^2}{\omega^2}-r\frac{\wpe^2}{\gamma_r \omega^2}
                \end{aligned} \label{eq:tensor}
        \end{equation},

        where $r= n_e/n_0$, $ \xi = k u_r / \Wce$, $r^\prime = 1-r$, $G_r =  \gamma_r-1/\gamma_r$, $\gamma_r = (1+u_r^2/c^2)^{1/2}$, $b=\gamma_r\omega-n\Wce$, and $J_n(\xi)$ refers to the usual Bessel function of order $n$.
        
        The following equation was solved to obtain both the frequency and growth rate of each extraordinary mode (X, Z, or R):
        \begin{equation} \label{eq:det}
                \varsigma_{xx}\varsigma_{yy}-\varsigma_{xy}\varsigma_{yx}=0   .
        \end{equation}         

        \subsection{ECMI of X and R with $n_e/n_0=24\%$} \label{sec:theory}

        To understand the PIC simulations, we first conducted linear theory analysis with the parameters of SKR sources reported by \cite{2011JGRA..11612222M}. We set $u_r = 0.18~c$, $\gamma_r = 1.016$, $n_e/n_0=24\%$, and $\wpe/\Wce = 0.046$.
        
\begin{figure}[!htb]
                
                \includegraphics[width=12cm]{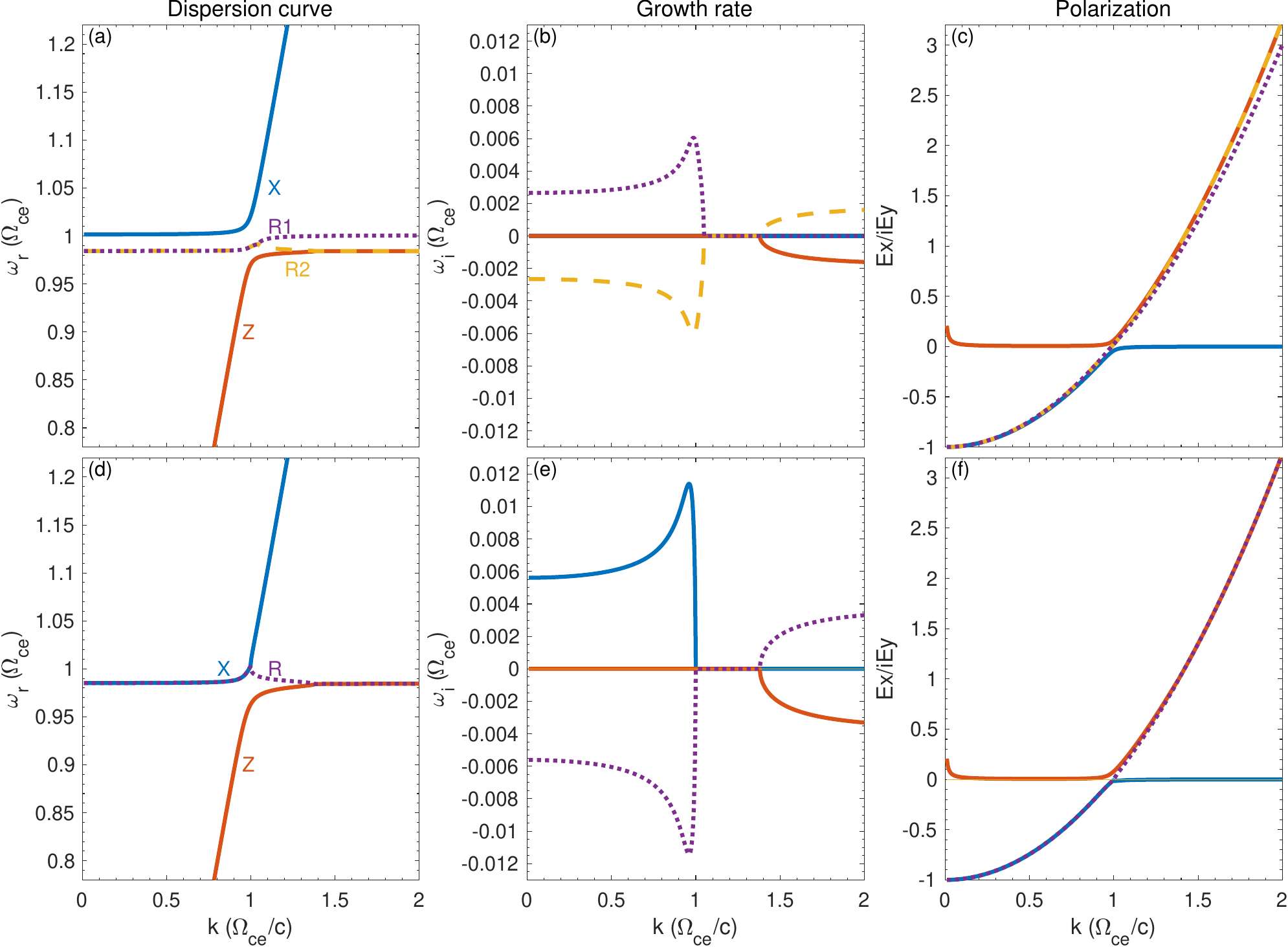}
                \caption{Dispersion relation (a, d) and growth rates (b, e) of the extraordinary wave modes according to the kinetic linear theory, with $n_e/n_0=24\%$ (upper panels) and 100\% (lower panels). The blue and red lines refer to the X and Z modes; the purple and yellow lines refer to the two branches of the relativistic modes (R1 and R2). Panels (c) and (f) present the corresponding intensity ratio of $E_x/iE_y$ for each mode.}
                \label{fig:A1}
        \end{figure}

        With a given $k$, we obtained four sets of complex roots $\omega=\omega_r + i\Gamma$ of Eq.~(\ref{eq:det}), representing different branches of the wave modes (see Fig.~\ref{fig:A1}(a--b)). The dispersion curves of the X mode lie in the range of $\omega_r>\Wce$ with a zero growth rate. The slow branch of the extraordinary mode (Z) is below $\Wce/\gamma_r$. 
        Within the range of $\Wce/\gamma_r<\omega_r<\Wce$ and $k\lesssim \Wce/c$, there exist two conjugate roots of the R mode (R1 and R2). The maximum absolute value of their growth rates is 0.006~$\Wce$. With increasing $k$ ($>\Wce/c$), R1 and R2 are decoupled and approach their respective  resonant frequency, that is, $\sim \Wce$ and $\Wce/\gamma_r$. 
                
        The perpendicular-propagating extraordinary waves are mainly carried by $E_x$ and $E_y$. The polarization was determined by 
        \begin{equation}
                \frac{E_x}{E_y} = -\frac{\varsigma_{xy}}{\varsigma_{xx}}. \label{eq:pol}
        \end{equation}
        Figure~\ref{fig:A1}(c) 
        displays the variations in $E_x/iE_y$ with $k$ for each extraordinary mode. 
        With increasing $k$, $E_x/iE_y$ rises from $-1$ to 0 for the X mode and from 0 to infinity for the Z mode. The $E_x/iE_y$ of R coincides with that of the X mode, with $k<\Wce/c$, and with that of the Z mode, with $k>\Wce/c$. 
        
        With $n_e/n_0=24\%$, R is the most unstable mode, with the highest growth rate at $k\sim\Wce/c$; where the three modes are basically linearly polarized with $E_x/iE_y\sim0$, the X mode cannot be amplified via the ECMI process.
        
        \subsection{Parameter studies on $n_e/n_0$, $\wpe/\Wce$, and $u_r$} \label{sec:ratio_theory}
        
\begin{figure*}[htb]
                \centering
                \includegraphics[width=14cm]{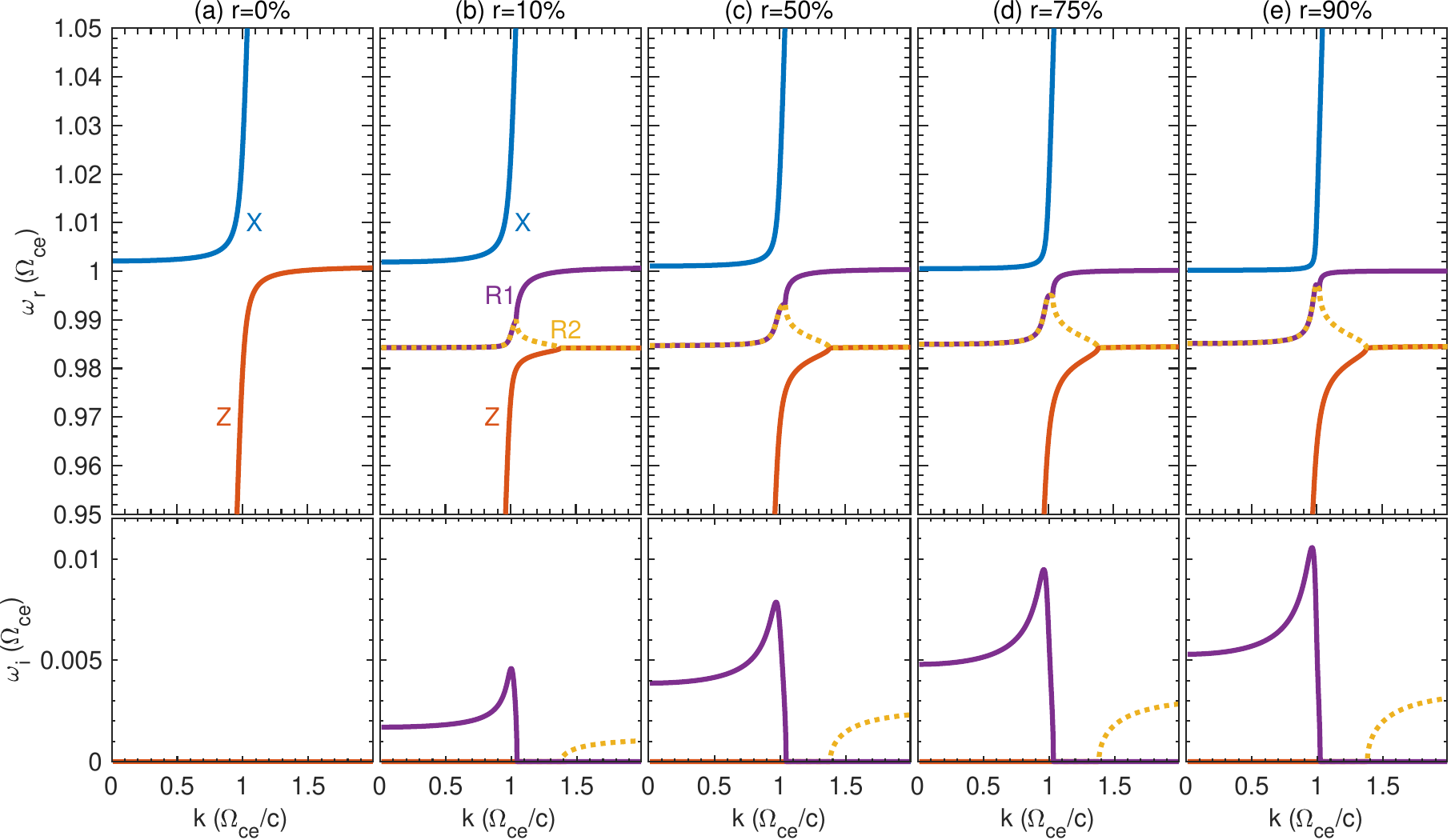}
                \caption{Dispersion relation (upper panels) and growth rates (lower panels) of the extraordinary wave modes (X, R1, R2, and Z) with varying $n_e/n_0$. }
                \label{fig:A2}
        \end{figure*}

        We varied $n_e/n_0$ from 0 to 100\%, keeping the other parameters fixed (Fig.~\ref{fig:A2}). With $n_e/n_0=0$, we obtained the magnetoionic cold-plasma dispersion relation with two branches (X and Z), with no instability.

        With $n_e/n_0>0$, we obtained conjugate pairs of solutions for the R1, R2, and Z modes,  and only those with positive growth rates are presented. The unstable R1 and R2 modes appear within $\Wce/\gamma<\omega<\Wce$ and their growth rates rise with $n_e/n_0$. With a higher abundance of energetic electrons, the X mode has a lower cutoff ($\omega_{\text{X}}$) and its dispersion curve becomes closer to that of the R mode. The X mode is always stable with $\omega_{\text{X}}>\Wce$ if $n_e/n_0<100\%$.
        
        With $n_e/n_0=100\%,$ the dispersion curve of the R and X modes merge together, with the cutoff $\omega_{\text{X}}<\Wce$. In this case, the X mode can be amplified via ECMI with the highest growth rate, $\sim 0.011~\Wce$. As Figure~\ref{fig:A1}(d--f) shows, the X mode is linearly polarized in $E_y$ ($E_x/iE_y\sim 0$) with $k\gtrsim\Wce/c$, and has the highest growth rate at $k\sim \Wce/c$. In summary, the linear theory suggests that the R mode is unstable if both cold and hot components are present, while the X mode can only be amplified with $n_e/n_0=100\%$.
        
        We further expanded the parameter study by varying $\wpe/\Wce$ and $u_r$ while fixing $n_e/n_0$ (=24\%). We find that the R mode is always dominant. With increasing $\wpe/\Wce$ (from 0.01 to 0.25) or $u_r$ (from 0.05 to 0.3~$c$), the dispersion curves of the R mode 
        appear in a broader range of frequencies with increasing growth rates.

        %
        

\end{appendix}

\end{document}